\documentclass[aps,prb,twocolumn,superscriptaddress,showpacs,showkeys]{revtex4-2}

\usepackage{color}
\usepackage{graphicx}
\usepackage{dcolumn}
\usepackage{bm}
\usepackage{float}
\usepackage{capt-of}
\usepackage[mathlines]{lineno}
\usepackage{footnote}
\usepackage{amsmath}
\usepackage{amsfonts}
\usepackage{mathptmx}
\usepackage{tabularx}
\usepackage{booktabs}
\usepackage{makecell}
\usepackage{multirow}
\usepackage{braket}
\usepackage{gensymb}
\usepackage{array}
\usepackage{svg}
\usepackage{siunitx}
\usepackage{placeins}
\usepackage{tikz}
\usetikzlibrary{calc}

\definecolor{legendblueone}{RGB}{8,48,107}
\definecolor{legendbluetwo}{RGB}{8,81,156}
\definecolor{legendbluethree}{RGB}{33,113,181}
\definecolor{legendbluefour}{RGB}{107,174,214}
\definecolor{legendredone}{RGB}{127,39,4}
\definecolor{legendredtwo}{RGB}{166,54,3}
\definecolor{legendredthree}{RGB}{230,85,13}
\definecolor{legendgreenone}{RGB}{0,109,44}
\definecolor{legendgreentwo}{RGB}{35,139,69}

\newcommand{\legendrule}[1]{%
  \textcolor{#1}{\rule[0.45ex]{8pt}{1.3pt}}%
}
\newcommand{\legendruleshort}[1]{%
  \textcolor{#1}{\rule[0.45ex]{3pt}{1.3pt}}%
}
\newcommand{\panellegendfont}{\fontsize{7.95pt}{7.95pt}\selectfont}
\newcommand{\batchlegendfont}{\fontsize{7.95pt}{7.95pt}\selectfont}
\newcommand{\latticelegendfont}{\fontsize{9pt}{9pt}\selectfont}

\newcommand{\legendConstraint}[1]{%
  \begin{scope}[
    overlay,
    shift={(img.north west)},
    x={($(img.north east)-(img.north west)$)},
    y={($(img.south west)-(img.north west)$)}
  ]
    \node[anchor=north west,inner sep=0,outer sep=0]
      at (0.2511544,0.0780000) {%
        \begingroup
        \panellegendfont
        \renewcommand{\arraystretch}{0.82}%
        \begin{tabular}{@{}c@{\hspace{2pt}}l@{}}
          \legendrule{legendblueone}   & $c_0=0.1$ \\
          \legendrule{legendbluetwo}   & $c_0=0.1$ \\
          \legendrule{legendbluethree} & $c_0=0.1$ \\
          \legendrule{legendredone}    & $c_0=0.5$ \\
          \legendrule{legendredtwo}    & $c_0=0.5$ \\
          \legendrule{legendredthree}  & $c_0=0.5$
        \end{tabular}%
        \endgroup
      };
  \end{scope}%
}

\newcommand{\legendWidth}[1]{%
  \begin{scope}[
    overlay,
    shift={(img.north west)},
    x={($(img.north east)-(img.north west)$)},
    y={($(img.south west)-(img.north west)$)}
  ]
    \node[anchor=north west,inner sep=0,outer sep=0]
      at (0.2820839,0.0780000) {%
        \begingroup
        \panellegendfont
        \renewcommand{\arraystretch}{0.82}%
        \begin{tabular}{@{}c@{\hspace{2pt}}l@{}}
          \legendrule{legendblueone}   & $d=32$ \\
          \legendrule{legendbluetwo}   & $d=32$ \\
          \legendrule{legendbluethree} & $d=32$ \\
          \legendrule{legendredone}    & $d=64$ \\
          \legendrule{legendredtwo}    & $d=64$ \\
          \legendrule{legendredthree}  & $d=64$
        \end{tabular}%
        \endgroup
      };
  \end{scope}%
}

\newcommand{\legendK}[1]{%
  \begin{scope}[
    overlay,
    shift={(img.north west)},
    x={($(img.north east)-(img.north west)$)},
    y={($(img.south west)-(img.north west)$)}
  ]
    \node[anchor=north west,inner sep=0,outer sep=0]
      at (0.2946416,0.0780000) {%
        \begingroup
        \panellegendfont
        \renewcommand{\arraystretch}{0.82}%
        \begin{tabular}{@{}c@{\hspace{2pt}}l@{}}
          \legendrule{legendbluetwo}   & $K=2$ \\
          \legendrule{legendbluethree} & $K=2$ \\
          \legendrule{legendredtwo}    & $K=4$ \\
          \legendrule{legendredthree}  & $K=4$ \\
          \legendrule{legendgreenone}  & $K=8$ \\
          \legendrule{legendgreentwo}  & $K=8$
        \end{tabular}%
        \endgroup
      };
  \end{scope}%
}

\newcommand{\legendBatch}[1]{%
  \begin{scope}[
    overlay,
    shift={(img.north west)},
    x={($(img.north east)-(img.north west)$)},
    y={($(img.south west)-(img.north west)$)}
  ]
    \node[anchor=north west,inner sep=0,outer sep=0]
      at (0.1950000,0.0800000) {%
        \begingroup
        \batchlegendfont
        \renewcommand{\arraystretch}{0.82}%
        \begin{tabular}{@{}c@{\hspace{1pt}}l@{}}
          \legendruleshort{legendbluetwo}   & $batch\!=\!1024$ \\
          \legendruleshort{legendbluethree} & $batch\!=\!1024$ \\
          \legendruleshort{legendredtwo}    & $batch\!=\!4096$ \\
          \legendruleshort{legendredthree}  & $batch\!=\!4096$
        \end{tabular}%
        \endgroup
      };
  \end{scope}%
}

\newcommand{\legendHalfFilled}[1]{%
  \begin{scope}[
    overlay,
    shift={(img.north west)},
    x={($(img.north east)-(img.north west)$)},
    y={($(img.south west)-(img.north west)$)}
  ]
    \node[anchor=north west,inner sep=0,outer sep=0]
      at (0.2762698,0.0780000) {%
        \begingroup
        \panellegendfont
        \renewcommand{\arraystretch}{0.82}%
        \begin{tabular}{@{}c@{\hspace{2pt}}l@{}}
          \legendrule{legendbluetwo}   & $d=32$ \\
          \legendrule{legendbluethree} & $d=32$ \\
          \legendrule{legendredtwo}    & $d=64$ \\
          \legendrule{legendredthree}  & $d=64$
        \end{tabular}%
        \endgroup
      };
  \end{scope}%
}

\newcommand{\panellegendoverlay}[1]{%
  \ifnum\pdfstrcmp{#1}{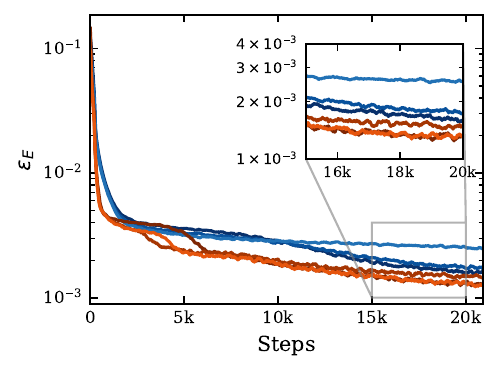}=0 \legendConstraint{#1}\fi
  \ifnum\pdfstrcmp{#1}{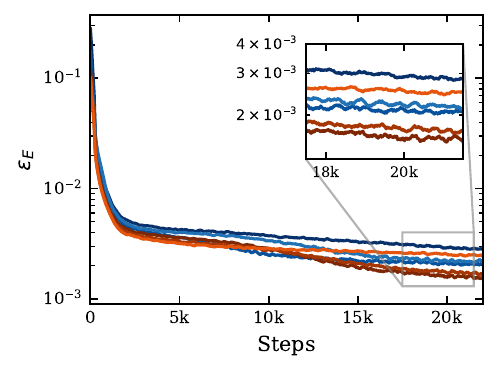}=0 \legendWidth{#1}\fi
  \ifnum\pdfstrcmp{#1}{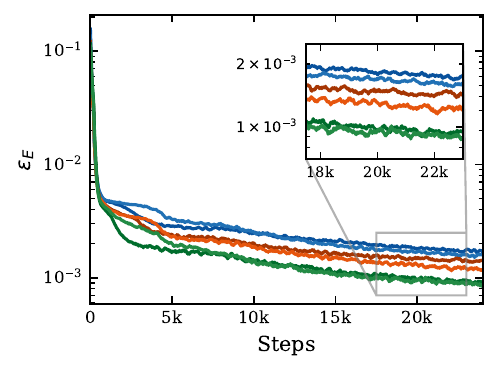}=0 \legendK{#1}\fi
  \ifnum\pdfstrcmp{#1}{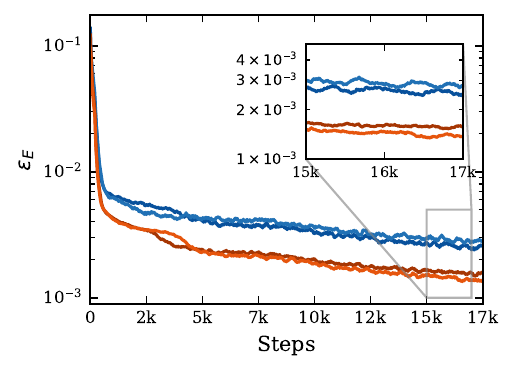}=0 \legendBatch{#1}\fi
  \ifnum\pdfstrcmp{#1}{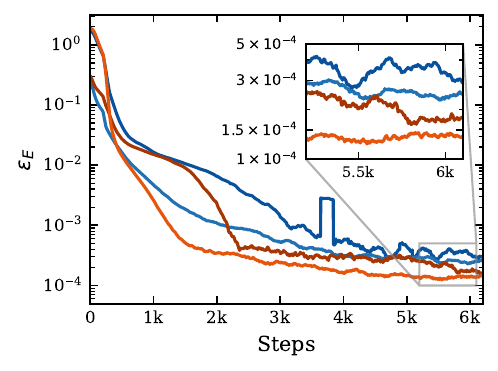}=0 \legendHalfFilled{#1}\fi
}

\newcommand{\panelplot}[2]{%
  \begin{tikzpicture}[baseline=(img.south)]
    \node[inner sep=0, outer sep=0] (img) {%
      \includegraphics[height=0.25322\textwidth]{#2}%
    };

    \panellegendoverlay{#2}%

    \path (img.south west) -- (img.north west)
      coordinate[pos=0.24] (labelpos);

    \node[
      overlay,
      anchor=east,
      xshift=50pt,
      font=\bfseries
    ] at (labelpos) {#1};
  \end{tikzpicture}%
}

\newcommand{\legendLattice}[1]{%
  \begin{scope}[
    overlay,
    shift={(img.north west)},
    x={($(img.north east)-(img.north west)$)},
    y={($(img.south west)-(img.north west)$)}
  ]
    \node[anchor=north west,inner sep=0,outer sep=0]
      at (0.3275474,0.0750000) {%
        \begingroup
        \latticelegendfont
        \renewcommand{\arraystretch}{0.82}%
        \begin{tabular}{@{}c@{\hspace{2pt}}l@{}}
          \legendrule{legendblueone}   & $d=128$ \\
          \legendrule{legendbluetwo}   & $d=128$ \\
          \legendrule{legendbluethree} & $d=128$ \\
          \legendrule{legendbluefour}  & $d=128$ \\
          \legendrule{legendredone}    & $d=64$ \\
          \legendrule{legendredtwo}    & $d=64$ \\
          \legendrule{legendredthree}  & $d=64$
        \end{tabular}%
        \endgroup
      };
  \end{scope}%
}

\newcommand{\latticeplot}[1]{%
  \begin{tikzpicture}[baseline=(img.south)]
    \node[inner sep=0,outer sep=0] (img) {%
      \includegraphics[width=0.90\columnwidth]{#1}%
    };
    \legendLattice{#1}%
  \end{tikzpicture}%
}

\usepackage[colorlinks,linkcolor=blue,anchorcolor=blue,citecolor=blue,urlcolor=blue,hyperindex,CJKbookmarks]{hyperref}

\newcolumntype{L}[1]{>{\raggedright\let\newline\\\arraybackslash\hspace{0pt}}m{#1}}
\newcolumntype{C}[1]{>{\centering\let\newline\\\arraybackslash\hspace{0pt}}m{#1}}
\newcolumntype{R}[1]{>{\raggedleft\let\newline\\\arraybackslash\hspace{0pt}}m{#1}}

\begin{document}

\title{Optimization dynamics of Transformer backflow neural quantum states for the two-dimensional Hubbard model}
\author{Zong-Yu Liao} \affiliation{School of Physics and Key Laboratory of Quantum State Construction and Manipulation (Ministry of Education), Renmin University of China, Beijing 100872, China} \author{Jia-Qi Wang}\affiliation{School of Physics and Key Laboratory of Quantum State Construction and Manipulation (Ministry of Education), Renmin University of China, Beijing 100872, China}\author{Rong-Qiang He} \email{rqhe@ruc.edu.cn} \affiliation{School of Physics and Key Laboratory of Quantum State Construction and Manipulation (Ministry of Education), Renmin University of China, Beijing 100872, China} \author{Zhong-Yi Lu} \email{zlu@ruc.edu.cn} \affiliation{School of Physics and Key Laboratory of Quantum State Construction and Manipulation (Ministry of Education), Renmin University of China, Beijing 100872, China} \affiliation{Hefei National Laboratory, Hefei 230088, China}
\date{\today}

\begin{abstract}
Building on the multi-determinant Transformer backflow neural quantum state (NQS) ansatz and the associated multi-stage training workflow for the doped two-dimensional Hubbard model, we investigate how the optimization dynamics of the NQS depend on several key optimization and architectural hyperparameters. The workflow consists of neural-network backflow (NNB) initialization, supervised Transformer pre-training, and main energy optimization using the Moment-Adaptive ReConfiguration Heuristic (MARCH) within variational Monte Carlo. Using the doped \(4\times4\) periodic Hubbard model at \(U=8\) as a baseline, we examine how the update-norm threshold, Transformer width, number of determinant channels, and Monte Carlo batch size affect convergence. We find that a moderate update constraint improves the efficiency of MARCH optimization, larger Transformer width and more determinant channels improve the expressive capacity of the ansatz, and larger Monte Carlo batches reduce sampling noise in the update direction. We further test the same workflow at half filling, weaker interaction strength, open boundary conditions, and on a larger \(8\times8\) doped lattice. These results identify practical optimization trends for Transformer backflow NQSs and highlight the balance between ansatz expressivity, MARCH update stability, and Monte Carlo sampling quality.
\end{abstract}

\maketitle

\section{Introduction}

The Hubbard model~\cite{annurev:/content/journals/10.1146/annurev-conmatphys-090921-033948,annurev:/content/journals/10.1146/annurev-conmatphys-031620-102024} is one of the central models for studying strongly correlated electrons. Despite its compact Hamiltonian, it contains a wide range of many-body phenomena, including Mott insulating behavior~\cite{RevModPhys.70.1039,PhysRevLett.108.076401,PhysRevB.95.235109,RevModPhys.68.13,PhysRevX.11.041013}, antiferromagnetic correlations~\cite{Mazurenko2017FermiHubbardAF,Cheuk2016SpinCharge2DFH,Parsons2016SpinCorrelationFH,PhysRevB.80.075116,PhysRevLett.124.017003,PhysRevB.69.085119}, charge and spin ordering tendencies~\cite{Zheng2017StripeOrder,PhysRevResearch.4.013239,PhysRevX.13.011007,PhysRevLett.104.116402,Mai2022IntertwinedSpinChargePair,Huang2018StripeOrderHubbard}, and the effects of carrier doping~\cite{Jiang2019DopedHubbardSC,PhysRevX.10.031016,PhysRevB.109.085121,scalapino2006numericalstudies2dhubbard}. In two dimensions, the Hubbard model has attracted sustained attention because of its possible connection to the microscopic physics of cuprate superconductors~\cite{Dong2022MechanismHubbardSC,PhysRevB.71.134527,PhysRevLett.85.1524,PhysRevLett.110.216405}. However, accurately obtaining its ground-state properties remains challenging. Existing numerical methods, including quantum Monte Carlo~\cite{PhysRevB.55.7464,PhysRevB.94.085103,PhysRevB.96.075156,PhysRevB.107.235124}, density matrix renormalization group (DMRG)~\cite{PhysRevLett.91.136403,PhysRevB.71.075108,PhysRevB.95.125125}, tensor-network methods~\cite{PhysRevB.93.045116,PhysRevB.107.165112,PhysRevB.100.195141,PhysRevLett.134.116502}, and embedding approaches~\cite{PhysRevLett.109.186404,PhysRevB.93.035126,Wu2019ProjectedDMET}, have provided important insights, but their performance can depend strongly on system size, boundary condition, doping level, and the presence of a fermionic sign problem.

Neural-network quantum states (NQSs) provide a complementary variational framework for quantum many-body problems. Since neural networks were introduced as variational wave-function ansatzes, a variety of architectures, including restricted Boltzmann machines~\cite{Carleo2017NQS,PhysRevB.96.205152,PhysRevX.8.011006}, convolutional neural networks~\cite{PhysRevB.109.245120,m4c7-qz8l}, recurrent networks, graph neural networks, and Transformers~\cite{2025arXiv250702644G,IbarraGarciaPadilla2025AutoNQS}, have been explored for representing many-body wave functions. Compared with conventional variational forms, NQSs can represent flexible correlation structures and can be optimized within variational Monte Carlo (VMC) using automatic differentiation. At the same time, practical NQS calculations are often limited not only by the expressive power of the ansatz, but also by the stability and efficiency of the optimization procedure~\cite{Chen2024EfficientOptimization,Rende2024LargeScaleSR}. In particular, local minima, slow convergence, stochastic sampling noise, and the difficulty of optimizing fermionic sign or phase structures can all affect the quality of the final variational state.

For fermionic lattice systems, these issues are especially important. The antisymmetric structure of the wave function must be represented properly, while the optimization must remain stable in the presence of Monte Carlo noise. A determinant-based construction provides a natural way to incorporate fermionic antisymmetry into the variational ansatz~\cite{RobledoMoreno2022HiddenFermion}. In a backflow determinant NQS~\cite{Luo2019Backflow}, the neural network generates configuration-dependent orbitals, and the wave-function amplitude is obtained from one or several determinants. This structure avoids treating the fermionic sign structure as a completely unconstrained neural-network output. Related neural-network-augmented Pfaffian constructions extend this orbital-based strategy to paired fermionic states and achieve high variational accuracy for both attractive and repulsive Hubbard models~\cite{Chen2025PfaffianNQS}. Instead, the remaining optimization problem is transferred to the configuration-dependent backflow orbitals and to the stability of the VMC training procedure.

Recent work~\cite{2025arXiv250702644G,Gu2026ParetoBackflow,Viteritti2026VariationalBias} has shown that Transformer-based NQSs combined with backflow orbital constructions can be applied to the doped two-dimensional Hubbard model, and that attention-based architectures provide a flexible way to encode correlations over different length scales. These developments suggest that Transformer NQSs are promising variational ansatzes for strongly correlated fermions. However, their practical performance depends on several coupled choices, including the expressive capacity of the network, the number of determinant channels, the Monte Carlo batch size, and the details of the optimization algorithm.

In this work, we study the optimization dynamics of a Transformer-based NQS for the two-dimensional Hubbard model. Rather than performing an exhaustive search over all possible training choices, we focus on identifying robust trends that control the VMC optimization. We use a multi-stage training strategy in which a lightweight neural-network backflow (NNB) wave function is first optimized, the Transformer is then pre-trained to reproduce the NNB-generated orbital matrices, and the resulting Transformer is finally optimized by the Moment-Adaptive ReConfiguration Heuristic (MARCH). Using the doped \(4\times4\) periodic boundary conditions (PBC) Hubbard model at \(U=8\) as a baseline, we examine the effects of the update-norm threshold, Transformer width, determinant number, and Monte Carlo batch size. We further test the same workflow at half filling, weaker interaction strength, open boundary conditions (OBC), and on a larger \(8\times8\) doped lattice. These calculations show that the optimization behavior is governed by the balance between ansatz expressivity, MARCH update stability, and sampling quality, and that the same training strategy can be applied across several Hubbard-model settings.

\section{Model and Method}
The variational ansatz and the multi-stage training workflow used in this work follow the Transformer backflow NQS framework introduced in Ref.~\cite{2025arXiv250702644G}, with the modifications and ablation settings specified below.

\subsection{The Hubbard model}

The Fermi-Hubbard model is the prototypical framework for investigating strongly correlated electronic systems and the emergence of unconventional superconductivity in cuprates. In this work, we consider the Hubbard model on a two-dimensional square lattice, governed by the following Hamiltonian:
\begin{equation}
\begin{split}
\hat{\mathcal{H}} = & -t \sum_{\langle i, j \rangle, \sigma} (c_{i\sigma}^\dagger c_{j\sigma} + \text{H.c.}) \\
    & - t' \sum_{\langle\langle i, j \rangle\rangle, \sigma} (c_{i\sigma}^\dagger c_{j\sigma} + \text{H.c.}) + U \sum_i n_{i\uparrow} n_{i\downarrow}
\end{split}
\end{equation}
where $c_{i\sigma}^\dagger$ ($c_{i\sigma}$) denotes the creation (annihilation) operator for an electron with spin $\sigma \in \{\uparrow, \downarrow\}$ at site $i$, and $n_{i\sigma} = c_{i\sigma}^\dagger c_{i\sigma}$ is the corresponding number operator. The parameter $t$ represents the hopping integral between nearest-neighbor sites $\langle i,j \rangle$, while $t'$ accounts for the hopping between next-nearest-neighbor sites $\langle\langle i,j \rangle\rangle$. The on-site Coulomb repulsion is quantified by $U$. The physical properties of the system are predominantly determined by the competition between the kinetic energy (governed by $t$ and $t'$) and the potential energy ($U$). In the strong correlation regime where $U/t \gg 1$, electrons tend to be localized, leading to a Mott insulating state at half-filling ($\langle n_i \rangle = 1$). The introduction of $t'$ is physically significant as it breaks the particle-hole symmetry of the square lattice. Specifically, a non-zero $t'$ can induce magnetic frustration and significantly alter the landscape of the fermionic sign problem, making the determination of the ground state even more challenging for conventional Monte Carlo methods. Our investigation focuses on a $4 \times 4$ cluster. We employ PBC to eliminate open-boundary edge effects and strictly preserve the translational symmetry of the lattice. We primarily examine the system at $t' = 0$ across various interaction strengths: the intermediate correlation regime ($U=4$), the strong correlation regime ($U=8$), and the deep Mott regime ($U=10$). The study encompasses both the half-filled case ($N_e = 16$) and the doped regime ($N_e = 14$, corresponding to $1/8$ hole doping). While the half-filled case at $t'=0$ remains sign-problem-free due to particle-hole symmetry, the doped regime presents a rigorous testbed for our Transformer-based NQS due to the presence of a severe fermionic sign problem.

\subsection{Transformer-based NQS}
To represent the many-body wave function $\Psi(n)$, we employ a NQS architecture based on the Transformer, which has demonstrated exceptional capabilities in capturing complex, non-local correlations in quantum many-body systems. 

Each physical site $i$ is treated as an independent “token" in the sequence. The four possible local states of the Hubbard model—empty ($|0\rangle$), spin-up ($|\uparrow\rangle$), spin-down ($|\downarrow\rangle$), and double-occupancy ($|\uparrow\downarrow\rangle$)—are mapped to one-hot vectors $\mathbf{v}_i \in \{0, 1\}^4$.

\begin{figure*}[!t]
\centering
\includegraphics[width=0.95\textwidth]{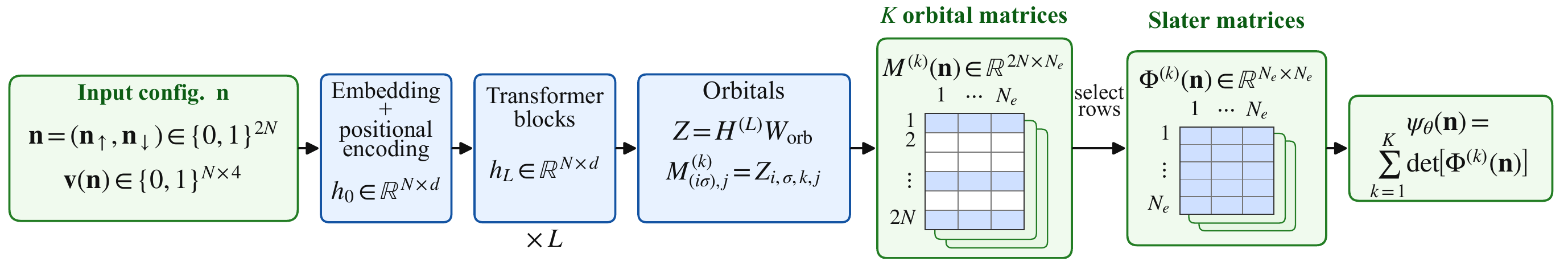}
\caption{
Schematic architecture of the main-training Transformer backflow NQS.
The input configuration is mapped to local one-hot tokens and processed by an embedding layer with positional encoding, followed by \(L\) Transformer blocks.
The orbital head produces \(K\) configuration-dependent orbital matrices \(M^{(k)}(n)\in\mathbb{R}^{2N\times N_e}\).
For each determinant channel, the occupied spin-orbital rows specified by \(I(n)\) are selected to form the Slater matrix \(\Phi^{(k)}(n)\in\mathbb{R}^{N_e\times N_e}\), and the final wave-function amplitude is obtained by summing the determinants over \(k\).
}
\label{fig:transformer_backflow_architecture}
\end{figure*}

The data flow begins at the embedding layer, which projects these sparse vectors into a high-dimensional continuous latent space. For each site $i$, the initial hidden representation $\mathbf{h}_i^{(0)} \in \mathbb{R}^d$ (where $d$ is the hidden dimension of the network) is constructed as:
\begin{equation}
    \mathbf{h}_i^{(0)} = \mathbf{W}_{\rm emb} \mathbf{v}_i + \mathbf{p}_i
\end{equation}
Here, $\mathbf{W}_{\rm emb} \in \mathbb{R}^{d \times 4}$ is a learnable projection matrix. The term $\mathbf{p}_i \in \mathbb{R}^d$ is a trainable positional encoding vector, which is essential for breaking the permutation invariance of the Transformer and injecting the two-dimensional spatial geometry of the lattice into the network. The resulting sequence matrix is denoted as $\mathbf{H}^{(0)} = [\mathbf{h}_1^{(0)}, \dots, \mathbf{h}_N^{(0)}]^T \in \mathbb{R}^{N \times d}$.

The embedded sequence $\mathbf{H}^{(0)}$ is then processed through $L$ identical Transformer blocks. The $l$-th block updates the hidden states $\mathbf{H}^{(l-1)}$ to $\mathbf{H}^{(l)}$ through two primary sub-layers: a Multi-Head Self-Attention (MHA) mechanism and a Feed-Forward Network (FFN).

To capture diverse physical correlation patterns (e.g., spin, charge, and long-range entanglement), the MHA employs $H$ independent attention heads. For a specific head $h$, the query, key, and value matrices are generated via linear transformations:
\begin{equation}
\begin{split}
    &\mathbf{Q}_h^{(l-1)} = \mathbf{H}^{(l-1)} \mathbf{W}_{Q,h}^{(l-1)}, \\ 
    &\mathbf{K}_h^{(l-1)} = \mathbf{H}^{(l-1)} \mathbf{W}_{K,h}^{(l-1)}, \\ 
    &\mathbf{V}_h^{(l-1)} = \mathbf{H}^{(l-1)} \mathbf{W}_{V,h}^{(l-1)}
\end{split}
\end{equation}
where $\mathbf{W}_{Q,h}^{(l-1)}, \mathbf{W}_{K,h}^{(l-1)}, \mathbf{W}_{V,h}^{(l-1)} \in \mathbb{R}^{d \times d_H}$ are learnable weight matrices, and $d_H = d/H$ is the dimension of each head. The output of head $h$, denoted as $\mathbf{H}_h \in \mathbb{R}^{N \times d_H}$, is computed as:
\begin{equation}
    \mathbf{H}_h^{(l-1)} = \text{softmax} \left( \frac{\mathbf{Q}_h^{(l-1)} {\mathbf{K}_h^{(l-1)}}^T}{\sqrt{d_H}} \right) \mathbf{V}_h^{(l-1)}
\end{equation}

As demonstrated in prior studies~\cite{2025arXiv250702644G}, the output of the attention heads explicitly learn to assign weights between distinct site pairs, thereby facilitating non-local information exchange and dynamically capturing the formation of specific physical orders, such as antiferromagnetic correlations or stripe patterns.

The outputs from all $H$ heads are concatenated and projected back to the original dimension $d$ using an output weight matrix $\mathbf{W}_O^{(l)} \in \mathbb{R}^{d \times d}$. Incorporating a residual connection, the intermediate hidden state $\mathbf{H}^{(l)\prime} \in \mathbb{R}^{N \times d}$ becomes:
\begin{equation}
    \mathbf{H}^{(l)\prime} = \mathbf{H}^{(l-1)} + [\mathbf{H}_1^{(l-1)}, \mathbf{H}_2^{(l-1)}, \dots, \mathbf{H}_H^{(l-1)}] \mathbf{W}_O^{(l-1)}
\end{equation}

Subsequently, the intermediate feature matrix $\mathbf{H}^{(l)\prime}$ is processed by a FFN. The transformation is given by:
\begin{equation}
    \text{FFN}^{(l)}(\mathbf{H}^{(l)\prime}) = \text{SiLU}(\mathbf{H}^{(l)\prime} \mathbf{W}^{(l)}_{\rm FFN})
\end{equation}
where $\mathbf{W}^{(l)}_{\rm FFN} \in \mathbb{R}^{d \times d}$ is a learnable weight matrix, and SiLU denotes the
sigmoid-weighted linear unit activation function,
$\operatorname{SiLU}(x)=x\,\operatorname{sigmoid}(x)$
\cite{Elfwing2018SiLU}. 

With a second residual connection, the final output of the $l$-th Transformer block is obtained:
\begin{equation}
    \mathbf{H}^{(l)} = \mathbf{H}^{(l)\prime} + \text{FFN}^{(l)}(\mathbf{H}^{(l)\prime})
\end{equation}

After passing through $L$ identical blocks, the final output matrix $\mathbf{H}^{(L)} \in \mathbb{R}^{N \times d}$ is obtained.

The fermionic sign problem fundamentally stems from the anti-symmetry of the many-body wave function under particle exchange, which leads to severe cancellations of positive and negative statistical weights during Monte Carlo sampling and exponentially increases the computational complexity in frustrated or doped regimes. To strictly enforce this fermionic anti-symmetry while maintaining high expressive power, our NQS architecture maps the latent features into a multi-determinant backflow representation.

To transform $\mathbf{H}^{(L)}$ into physical wave function components, we apply a spin-independent linear projection layer:
\begin{equation}
    \mathbf{Z} = \mathbf{H}^{(L)} \mathbf{W}_{\rm orb}
\end{equation}
where the learnable weight matrix $\mathbf{W}_{\rm orb} \in \mathbb{R}^{d \times (2 \cdot K \cdot N_e)}$.

For a given input lattice configuration
$\mathbf{n} = (\mathbf{n}_{\uparrow}, \mathbf{n}_{\downarrow}) \in \{0, 1\}^{2N}$,
the output $\mathbf{Z}$ is reorganized into $K$ sets of orbital
matrices $\mathbf{M}^{(k)} \in \mathbb{R}^{2N\times N_e}$ for
$k=1,\ldots,K$. The row dimension $2N$ corresponds to the spin-orbital degrees of freedom (accounting for both spin-up and spin-down possibilities at $N$ spatial sites), while the column dimension corresponds to the $N_e$ electrons. Here, the matrices $\mathbf{M}^{(k)}$ represent fermionic backflow orbitals. Unlike static single-particle orbitals in conventional mean-field theory, each orbital value $M^{(k)}_{m, j}$ depends non-linearly on the entire many-body configuration $n$ processed by the Transformer. This context-aware property allows the orbitals to capture higher-order electronic correlations.

In order to construct the Slater matrices, the network then identifies the global indices of all occupied states to form an index set $I(n) = \{s_1, s_2, \dots, s_{N_e}\}$, where $s_m \in \{1, \dots, 2N\}$.

For the $k$-th orbital set, the corresponding $N_e \times N_e$ Slater matrix $\Phi^{(k)}(n)$ is constructed by dynamically gathering the rows from $\mathbf{M}^{(k)}$ that match the occupied indices $I(n)$:
\begin{equation}
    \Phi^{(k)}_{m,j}(n) = M^{(k)}_{s_m, j}
\end{equation}

To compute the final variational wave function amplitude $\psi_{\theta}(n)$, we evaluate the sum of the determinants of these matrices $K$:
\begin{equation}
    \psi_{\theta}(n) = \sum_{k=1}^K \det \left[ \Phi^{(k)}(n) \right]
\end{equation}

The main Transformer backflow architecture used in the VMC optimization is summarized in Fig.~\ref{fig:transformer_backflow_architecture}.

\subsection{Multi-training-stage Strategy}

Directly minimizing the variational energy of a highly expressive deep Transformer network from a random parameter initialization poses severe numerical challenges. To circumvent this, we implement a multi-stage training protocol designed to capture a coarse physical picture before engaging the full capacity of the Transformer. This protocol consists of three sequential phases: (i) a computationally lightweight neural network backflow is first optimized; (ii) the Transformer ansatz is subsequently pre-trained in a supervised manner to reproduce the backflow orbital matrices generated by the optimized lightweight NNB on sampled configurations; and (iii) the pre-trained Transformer undergoes the main VMC optimization. This progressive strategy provides a physically meaningful initial state $|\psi_{\theta_0}\rangle$, thus stabilizing the subsequent optimization dynamics.

\subsubsection{Neural Network Backflow Architecture}

To provide a computationally efficient initialization, we construct a NNB architecture. This NNB serves as the first stage of our training protocol. It maps the real-space electronic configuration to a set of spin-dependent backflow orbitals, which are subsequently used to evaluate the wave function amplitude via a Slater determinant.

Given a spin configuration $\mathbf{n} = (\mathbf{n}_{\uparrow}, \mathbf{n}_{\downarrow}) \in \{0, 1\}^{2N}$, the NNB employs two independent Multi-Layer Perceptrons (MLPs) to generate the orbitals for spin-up and spin-down electrons. To explicitly capture the correlation between opposite spins, both MLPs take the full configuration $\mathbf{n}$ as input. The intermediate features is given by a two-hidden-layer feed-forward structure:
\begin{equation}
\begin{split}
&\tilde{M}_{\sigma}(\mathbf{n}) = \\&W^{(3)}_{\sigma} \text{ReLU}\left(W^{(2)}_{\sigma} \text{ReLU}\left(W^{(1)}_{\sigma} \mathbf{n} + \mathbf{b}^{(1)}_{\sigma}\right) + \mathbf{b}^{(2)}_{\sigma}\right) + \mathbf{b}^{(3)}_{\sigma}
\end{split}
\end{equation}
for the spin index $\sigma \in \{\uparrow, \downarrow\}$. Here, $W^{(1)}_{\sigma} \in \mathbb{R}^{d_{\text{hid}} \times 2N}$, $W^{(2)}_{\sigma} \in \mathbb{R}^{d_{\text{hid}} \times d_{\text{hid}}}$, and $W^{(3)}_{\sigma} \in \mathbb{R}^{(N \cdot N_e/2) \times d_{\text{hid}}}$ are learnable weight matrices, with $\mathbf{b}^{(i)}_{\sigma}$ being the corresponding bias vectors. The parameter $d_{\text{hid}}$ denotes the hidden dimension of the network, and $N_e$ is the total number of electrons in the system.

The output vector $\tilde{M}_{\sigma}(\mathbf{n})$ is then reshaped into an orbital matrix $M_{\sigma}(\mathbf{n}) \in \mathbb{R}^{N \times (N_e/2)}$. The $i$-th row of $M_{\sigma}(\mathbf{n})$ represents the orbital evaluated at the spatial site $i$. To construct the physical wave function, we extract the occupied orbitals to form the Slater matrix. Let $I_{\sigma}(\mathbf{n}) = \{ i \mid n_{i,\sigma} = 1 \}$ be the index set of occupied sites for spin $\sigma$. We construct the block Slater matrix $\Phi_{\sigma}(\mathbf{n}) \in \mathbb{R}^{(N_e/2) \times (N_e/2)}$ by gathering the rows of $M_{\sigma}(\mathbf{n})$ that match the occupied indices:
\begin{equation}
[\Phi_{\sigma}(\mathbf{n})]_{m, j} = [M_{\sigma}(\mathbf{n})]_{s_m, j}, \quad s_m \in I_{\sigma}(\mathbf{n})
\end{equation}
where $m, j \in \{1, \dots, N_e/2\}$. The global Slater matrix for the system is then assembled as a block-diagonal matrix:
\begin{equation}
\Phi(\mathbf{n}) = 
\begin{pmatrix} 
\Phi_{\uparrow}(\mathbf{n}) & \mathbf{0} \\ 
\mathbf{0} & \Phi_{\downarrow}(\mathbf{n}) 
\end{pmatrix}
\end{equation}
Finally, the variational wave function amplitude of the NNB is computed as the determinant of this block-diagonal matrix:
\begin{equation}
\psi_{\text{NNB}}(\mathbf{n}) = \det[\Phi(\mathbf{n})] = \det[\Phi_{\uparrow}(\mathbf{n})] \det[\Phi_{\downarrow}(\mathbf{n})]
\end{equation}

The NNB parameters are optimized by minimizing the variational energy within VMC. At each optimization step, electronic configurations are sampled from the probability distribution defined by the current NNB wave function,
\begin{equation}
p_{\rm NNB}(n)
=
\frac{|\psi_{\rm NNB}(n)|^2}
{\sum_n |\psi_{\rm NNB}(n)|^2}.
\end{equation}
The variational energy is estimated as the Monte Carlo average of the local energy,
\begin{equation}
E_{\rm NNB}
=
\left\langle
E_{\rm loc}^{\rm NNB}(n)
\right\rangle_{p_{\rm NNB}(n)},
\end{equation}
where
\begin{equation}
E_{\rm loc}^{\rm NNB}(n)
=
\sum_{n'}
\frac{\psi_{\rm NNB}(n')}{\psi_{\rm NNB}(n)}
\langle n'|\hat{H}|n\rangle .
\end{equation}
This energy is minimized using the Adam optimizer~\cite{kingma2014adam}.

By utilizing this single-determinant ($K=1$) structure composed of shallow fully-connected layers, the NNB maintains a relatively small parameter space. It acts as an inexpensive structural prior to rapidly approximate the ground state energy landscape, preparing physical configurations for the subsequent training of the target Transformer NQS.

\subsubsection{Pre-training}

After the NNB wave function has been optimized, we use it to initialize the Transformer ansatz through a supervised pre-training stage. The purpose of this stage is to transfer the coarse physical structure learned by the lightweight NNB to the more expressive Transformer representation. 

In this stage, the optimized NNB is used as the sampling wave function. Starting from thermalized configurations, we perform Markov-chain updates with the NNB parameters fixed and collect a set of electronic configurations
\begin{equation}
\mathcal{D}_{\rm cfg}=\{n_i\}_{i=1}^{N_{\rm pre}},
\end{equation}
where each configuration is represented in the binary occupation form
\begin{equation}
n_i=(n_{i,\uparrow},n_{i,\downarrow})\in\{0,1\}^{2N}.
\end{equation}
For each sampled configuration \(n_i\), the optimized NNB produces a block-diagonal orbital matrix
\begin{equation}
M_{\rm NNB}(n_i)\in \mathbb{R}^{2N\times N_e},
\end{equation}
which is used as the supervised target. The resulting pre-training data can therefore be written as
\begin{equation}
\mathcal{S}_{\rm pre}
=
\left\{
n_i,\,
M_{\rm NNB}(n_i)
\right\}_{i=1}^{N_{\rm pre}} .
\end{equation}

For the Transformer input, the same configuration \(n_i\) is converted into the local occupation representation introduced above. At each physical site \(j\), the local state \(\sigma_j\) takes one of the four possible values: empty, spin-up, spin-down, or double occupancy. It is then mapped to a one-hot vector \(v_j\in\{0,1\}^4\) and fed into the Transformer sequence. The Transformer outputs \(K\) sets of backflow orbital matrices,
\begin{equation}
\left\{
M_\theta^{(k)}(n_i)
\right\}_{k=1}^{K},
\qquad
M_\theta^{(k)}(n_i)\in\mathbb{R}^{2N\times N_e}.
\end{equation}

During pre-training, the Transformer parameters are trained to reproduce the NNB-generated orbital matrices. Since the NNB provides a single orbital matrix while the Transformer contains \(K\) determinant channels, the same NNB target is matched by each channel. The supervised loss is taken as
\begin{equation}
\mathcal{L}_{\rm pre}(\theta)
=
\frac{1}{N_{\rm pre}K}
\sum_{i=1}^{N_{\rm pre}}
\sum_{k=1}^{K}
\left\|
M_\theta^{(k)}(n_i)
-
M_{\rm NNB}(n_i)
\right\|_F^2 .
\end{equation}
This loss is minimized using the Adam optimizer.

After this supervised stage, the resulting Transformer parameters are used as the initial parameters for the main VMC optimization. 

\subsubsection{Main training}

The main optimization is performed using MARCH. Let \(N_p\) be the number of real variational parameters, and let \(\boldsymbol{\theta}\in\mathbb{R}^{N_p}\) denote the flattened parameter vector. For each configuration \(n\), we define the logarithmic derivative
\begin{equation}
O_\alpha(n)
=
\frac{\partial \log \psi_\theta(n)}
{\partial \theta_\alpha},
\qquad
\alpha=1,\ldots,N_p .
\end{equation}
For a Monte Carlo batch \(\{n_i\}_{i=1}^{B}\), the centered logarithmic derivatives are collected into the matrix
\begin{equation}
\widetilde O_{i\alpha}
=
O_\alpha(n_i)
-
\frac{1}{B}
\sum_{j=1}^{B}
O_\alpha(n_j),
\qquad
\widetilde O\in\mathbb{C}^{B\times N_p}.
\end{equation}
The corresponding imaginary-time target vector is defined as
\begin{equation}
\widetilde\epsilon_i
=
-\delta\tau
\left[
E_{\rm loc}^{\theta}(n_i)-E_\theta
\right],
\qquad
\widetilde{\boldsymbol\epsilon}\in\mathbb{C}^{B},
\end{equation}
where \(\delta\tau\) is the imaginary-time step. 

 At optimization step \(k\), the MARCH direction \(d\boldsymbol{\theta}_k\in\mathbb{R}^{N_p}\) is defined as
\begin{equation}
d\boldsymbol{\theta}_{k}
=
\operatorname*{arg\,min}_{d\boldsymbol{\theta}'\in \mathbb{R}^{N_p}}
\frac{1}{\lambda}
\left\|
\widetilde O d\boldsymbol{\theta}'
-
\widetilde{\boldsymbol\epsilon}
\right\|_2^2
+
\left\|
D_{k-1}
\left(
d\boldsymbol{\theta}'
-
\boldsymbol\phi_{k-1}
\right)
\right\|_2^2 ,
\end{equation}
where
\begin{equation}
D_{k-1}
=
\operatorname{diag}
\left[
(\boldsymbol v_{k-1}+\varepsilon\mathbf{1}_{N_p})^{1/4}
\right]
\in\mathbb{R}^{N_p\times N_p}.
\end{equation}
Here, \(\boldsymbol\phi_{k-1}\in\mathbb{R}^{N_p}\) is the first-moment estimator, which shifts the update direction toward the accumulated momentum direction, while \(\boldsymbol v_{k-1}\in\mathbb{R}^{N_p}\) is the second-moment estimator, which rescales different parameter directions according to their recent fluctuations. The small positive constant \(\varepsilon\) is introduced for numerical stability.

After \(d\boldsymbol{\theta}_k\) is obtained, the moment estimators are updated as
\begin{equation}
\boldsymbol\phi_k
=
\mu d\boldsymbol{\theta}_k ,
\end{equation}
and
\begin{equation}
\boldsymbol v_k
=
\beta\boldsymbol v_{k-1}
+
\left(
d\boldsymbol{\theta}_k
-
d\boldsymbol{\theta}_{k-1}
\right)^{\odot 2},
\end{equation}
where \(\mu\) and \(\beta\) are the first- and second-moment parameters, respectively, and \(\odot 2\) denotes element-wise square. In this way, MARCH combines the geometry-aware update of stochastic reconfiguration~\cite{10.1063/1.2746035,PhysRevLett.87.043401,PhysRevB.64.024512} with an adaptive-momentum mechanism: stable directions can be followed more efficiently, while directions with large recent fluctuations are suppressed.

To avoid rare but destructive updates caused by large fluctuations of \(E_{\rm loc}^{\theta}\), we additionally apply a time-dependent norm constraint to the MARCH direction,
\begin{equation}
\Delta\boldsymbol{\theta}_k
=
\min\left(
1,\,
\frac{c_k}{\|d\boldsymbol{\theta}_k\|_2+\varepsilon}
\right)
d\boldsymbol{\theta}_k ,
\end{equation}
with
\begin{equation}
c_k
=
c_0
\left[
1+
\frac{\max(0,k-k_0)}{k_0}
\right]^{-1}.
\end{equation}
This constraint leaves ordinary updates unchanged and only rescales exceptionally large steps, thereby preventing sudden parameter jumps and stabilizing the optimization trajectory.

The Transformer parameters are finally updated as
\begin{equation}
\boldsymbol{\theta}_{k+1}
=
\boldsymbol{\theta}_{k}
+
\Delta\boldsymbol{\theta}_k .
\end{equation}

Local-energy clipping is employed in both the VMC training of
the NNB and the main VMC optimization of the Transformer. For memory-efficient evaluation, each Monte Carlo batch
 $B$ is partitioned
into $N_{\mu}=B/B_{\mu}$ disjoint microbatches, each of size
$B_{\mu}=64$. For each microbatch,
we compute the mean $m_{\mu}$ and the mean absolute deviation
$a_{\mu}$ of the real parts of the raw local energies,
\begin{equation}
\begin{aligned}
m_{\mu}
&=
\frac{1}{N_{\mu}}
\sum_{i\in N_{\mu}}
\operatorname{Re}
\left[
E_{\mathrm{loc},i}^{\mathrm{raw}}
\right],
\\
a_{\mu}
&=
\frac{1}{N_{\mu}}
\sum_{i\in {N}_{\mu}}
\left|
\operatorname{Re}
\left[
E_{\mathrm{loc},i}^{\mathrm{raw}}
\right]
-
m_{\mu}
\right|.
\end{aligned}
\end{equation}
The clipped local energy is defined as
\begin{equation}
\begin{aligned}
\widetilde{E}_{\mathrm{loc},i}
={}&
\operatorname{clip}
\left(
\operatorname{Re}
\left[
E_{\mathrm{loc},i}^{\mathrm{raw}}
\right],
m_{\mu}-\kappa_{\mathrm{clip}}a_{\mu},
m_{\mu}+\kappa_{\mathrm{clip}}a_{\mu}
\right)
\\
&+
\mathrm{i}\,
\operatorname{Im}
\left[
E_{\mathrm{loc},i}^{\mathrm{raw}}
\right],
\end{aligned}
\end{equation}
where $\kappa_{\mathrm{clip}}$ = 5 is the clipping scale and
$\operatorname{clip}(x,l,u)=\min[\max(x,l),u]$. The
construction of the clipping interval follows the
mean-centered mean-absolute-deviation prescription introduced
in the original FermiNet work~\cite{Pfau2020FermiNet}, with
the adaptations described above for the present complex-valued
lattice wave function and microbatch implementation. Thus, only
the real part of the local energy is clipped, while the
imaginary part remains unchanged. This procedure suppresses the disproportionate
contribution of local-energy outliers before the optimization
direction is determined. 

\section{Results}
\subsection{Error metric and numerical setup}

We measure the optimization accuracy by the relative energy error
\begin{equation}
\epsilon_E(r)
=
\frac{
\left|E(r)-E_{\rm ref}\right|
}{
\left|E_{\rm ref}\right|
},
\end{equation}
where \(r\) is the training step, \(E(r)\) is the VMC energy at step \(r\), and \(E_{\rm ref}\) is the reference ground-state energy for the corresponding system. For the main \(4\times4\) doped PBC benchmark at \(U=8\) and \(N_e=14\), we use \(E_{\rm ref}=-11.86884\), which is from exact diagonalization. The reference values for the other physical settings are taken from the corresponding benchmark or DMRG results~\cite{r4q9-4yvj} and are used consistently when computing \(\epsilon_E\).

Unless otherwise specified, the hyperparameter ablations are performed on the \(4\times4\) Hubbard model with PBC, and \(1/8\) hole doping. 

The baseline hyperparameters used in the ablation study are summarized in Table~\ref{tab:baseline_hyperparams}. In each ablation, only the parameter under study is varied, while the remaining parameters are kept fixed.

\begin{table}[t]
\centering
\caption{Baseline hyperparameters for the \(4\times4\), \(U=8\), \(N_e=14\) PBC ablation experiments.}
\label{tab:baseline_hyperparams}
\begin{ruledtabular}
\begin{tabular}{cc}
Parameter & Baseline value \\
\hline
Monte Carlo batch size \(B\) & \(4096\) \\
Transformer hidden dimension \(d\) & \(64\) \\
Number of determinant channels \(K\) & \(4\) \\
Initial norm threshold \(c_0\) & \(0.5\) \\
MARCH regularization \(\lambda\) & \(0.001\) \\
Norm-threshold decay scale \(k_0\) & \(8000\) \\
\end{tabular}
\end{ruledtabular}
\end{table}

The NNB and supervised Transformer pre-training stages are used only to provide an initialization for the main VMC optimization. Their schedules are kept fixed within each ablation group. The NNB hidden dimension is \(d_{\rm hid}=256\) for the baseline \(4\times4\), \(U=8\), PBC ablations, and \(d_{\rm hid}=64\) for the \(U=4\), OBC, and \(8\times8\) tests.

\onecolumngrid
\begingroup
\vspace*{8pt}
\setlength{\abovecaptionskip}{0pt}
\centering
\setlength{\tabcolsep}{0pt}
\renewcommand{\arraystretch}{1.0}

\resizebox{14cm}{!}{%
\begin{tabular}{@{}ll@{}}
\panelplot{(a)}{fig_constraint_ablation_legend_clean.pdf}
&
\panelplot{(b)}{fig_width_ablation_legend_clean.pdf}
\\[-0.2em]
\panelplot{(c)}{fig_K_ablation_legend_clean.pdf}
&
\panelplot{(d)}{fig_batch_ablation_legend_clean.pdf}
\end{tabular}%
}

\captionof{figure}{Baseline ablation results for the doped \(4\times4\) PBC Hubbard model at \(U=8\) and \(N_e=14\). The four panels compare (a) the initial update-norm threshold \(c_0\), (b) the Transformer hidden dimension \(d\), (c) the number of determinant channels \(K\), and (d) the Monte Carlo batch size \(B\). In each panel, only the indicated parameter is varied, while the remaining main-training settings are fixed for that comparison. In panel (a), the less restrictive choice \(c_0=0.5\) reaches the \(10^{-3}\) error level more efficiently, whereas \(c_0=0.1\) remains at errors of a few \(10^{-3}\). In panel (b), increasing the width from \(d=32\) to \(d=64\) lowers the late-time error from the several-\(10^{-3}\) range toward the \(10^{-3}\) level. In panel (c), increasing the number of determinant channels systematically improves the late-time accuracy, with \(K=8\) giving the lowest errors, close to \(10^{-3}\). In panel (d), increasing the batch size from \(B=1024\) to \(B=4096\) reduces the late-time error from a few \(10^{-3}\) to around the \(10^{-3}\) level and also suppresses visible Monte Carlo fluctuations.}
\label{fig:ablation_summary}
\endgroup
\newpage
\twocolumngrid

\begin{figure*}[!t]
\centering
\setlength{\tabcolsep}{0pt}
\renewcommand{\arraystretch}{1.0}

\resizebox{14cm}{!}{%
\begin{tabular}{@{}ll@{}}
\panelplot{(a)}{fig_half_filled_pbc_legend_clean.pdf}
&
\panelplot{(b)}{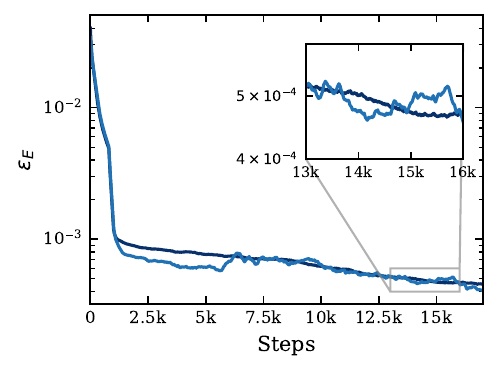}
\\[-0.2em]
\panelplot{(c)}{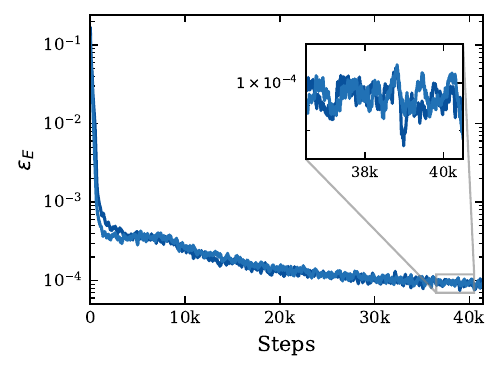}
&
\panelplot{(d)}{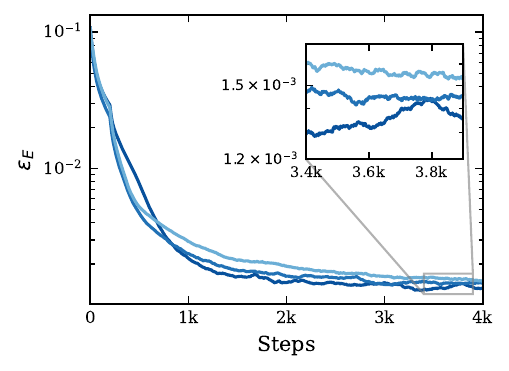}
\end{tabular}%
}
\caption{Optimization behavior across different \(4\times4\) Hubbard-model settings. The panels show (a) the half-filled PBC system at \(U=8\), (b) the doped PBC system at weaker interaction \(U=4\), (c) the half-filled OBC system at \(U=8\), and (d) the doped OBC system at \(U=8\). Panel (a) compares \(d=32\) and \(d=64\) with \(K=4\), \(B=4096\), and \(c_0=0.1\), making it directly comparable to the width ablation of the strongly correlated doped \(U=8\) baseline in Fig.~\ref{fig:ablation_summary}(b). The half-filled PBC system in panel (a) reaches the \(10^{-4}\) error level, whereas the corresponding doped \(U=8\) baseline in Fig.~\ref{fig:ablation_summary}(b) remains mainly at the \(10^{-3}\) level. Panel (b) uses the same main Transformer/MARCH settings as the \(c_0=0.5\) baseline curves in Fig.~\ref{fig:ablation_summary}(a), namely \(d=64\), \(K=4\), \(B=4096\), and \(c_0=0.5\), and shows that reducing the interaction to \(U=4\) brings the doped system down to the \(10^{-4}\) error level. Panels (c) and (d) compare the two OBC systems under the same main-training setup, \(d=64\), \(K=4\), \(B=4096\), and \(c_0=0.5\): the half-filled OBC case in panel (c) reaches the \(10^{-4}\) level, while the doped OBC case in panel (d) remains closer to the \(10^{-3}\) level. }
\label{fig:physical_settings_4x4}
\end{figure*}

\begin{figure}[!t]
\centering
\latticeplot{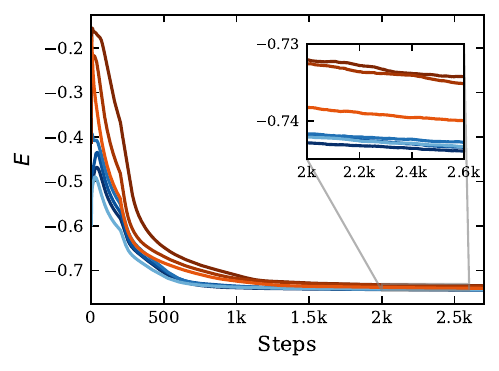}
\caption{Optimization on the larger doped \(8\times8\) PBC Hubbard lattice at \(U=8\) and \(N_e=56\).
Because a strict reference energy is not used for this larger system in the present comparison, we plot the variational energy per site, denoted by \(E\) in this figure.
The calculation uses \(K=4\), \(B=1024\), and \(c_0=0.5\), and compares Transformer widths \(d=64\) and \(d=128\).
Both widths lead to stable optimization on the larger lattice, with the energy per site decreasing to the \(E\simeq -0.73\)--\(-0.74\) range.
The wider Transformer reaches a lower late-time energy, with a visible separation from the \(d=64\) curves on the \(10^{-3}\)--\(10^{-2}\) energy-per-site scale, indicating that increased model capacity remains beneficial in the larger doped system.}
\label{fig:l8_pbc}
\end{figure}

For the systems with available benchmark energies, we plot the relative energy error \(\epsilon_E\). For the larger \(8\times8\) calculation, where we focus on the optimization behavior rather than a strict comparison to an exact reference energy, we instead plot the variational energy per site, denoted as \(E\) in Fig.~\ref{fig:l8_pbc}. Unless otherwise stated, the curves are smoothed by a moving average over 200 training steps. The \(U=4\), \(4\times4\), doped PBC result is smoothed using a wider 800-step moving average to suppress the larger Monte Carlo fluctuations in the absolute energy estimates. Different shades of the same color denote independent random seeds with the same physical and hyperparameter setting.

\subsection{A moderate update-norm threshold improves optimization efficiency}

Figure~\ref{fig:ablation_summary}(a) compares the optimization trajectories obtained with two initial norm thresholds, \(c_0=0.1\) and \(c_0=0.5\). The smaller threshold produces a more conservative update by damping the MARCH direction more strongly. However, this also restricts the effective parameter displacement during the main VMC optimization, leading to a slower reduction of \(\epsilon_E\).

By contrast, the larger threshold \(c_0=0.5\) allows the update direction to be followed more effectively while still suppressing exceptionally large steps. The resulting trajectory decreases faster and reaches a lower error within the plotted training window. This shows that the norm constraint controls not only numerical stability, but also the practical efficiency of the optimization. In the present setting, \(c_0=0.5\) provides a better balance between stability and update efficiency than the more restrictive choice \(c_0=0.1\).

\subsection{Increasing the Transformer width improves convergence accuracy}

Figure~\ref{fig:ablation_summary}(b) compares two Transformer widths, \(d=32\) and \(d=64\), under otherwise identical training conditions. Increasing the hidden dimension leads to a clear improvement in the optimization trajectory. The wider Transformer reaches a lower energy error and shows more favorable late-time convergence across independent random seeds.

This behavior reflects the role of the hidden representation in the Transformer backflow ansatz. Before the orbital matrices are generated, the self-attention layers must encode the nonlocal correlations contained in the electronic configuration. A larger hidden dimension provides more capacity for representing these correlations and for constructing more flexible configuration-dependent backflow orbitals.

\subsection{More determinant channels enhance the expressive power of the ansatz}

Figure~\ref{fig:ablation_summary}(c) compares the optimization trajectories obtained with \(K=2\), \(K=4\), and \(K=8\) determinant channels. Increasing \(K\) systematically improves the late-time accuracy. The \(K=2\) runs remain at higher energy errors, while increasing the number of channels leads to lower variational errors within the plotted training window.

This trend reflects the role of the determinant expansion in the Transformer backflow ansatz. The variational wave function is written as
\begin{equation}
\psi_\theta(n)
=
\sum_{k=1}^{K}
\det\left[\Phi^{(k)}(n)\right],
\end{equation}
where each determinant channel provides a distinct set of configuration-dependent backflow orbitals. A larger \(K\) allows the ansatz to combine multiple fermionic structures instead of relying on a small number of determinant components. This additional flexibility is especially important in the doped regime, where the correlated ground state is difficult to represent with a single dominant determinant. The improvement observed as \(K\) is increased therefore indicates that the multi-determinant structure is a central source of expressive power in the present Transformer NQS.

\subsection{Larger Monte Carlo batches reduce sampling noise and improve convergence}

Figure~\ref{fig:ablation_summary}(d) compares the optimization trajectories obtained with two Monte Carlo batch sizes, \(B=1024\) and \(B=4096\). The larger batch produces a smoother trajectory and reaches a lower late-time error within the plotted training window.

This behavior originates from the stochastic nature of the VMC estimates entering the MARCH update. Both the local energy and the logarithmic derivatives of the wave function are evaluated from Monte Carlo samples. A smaller batch therefore gives noisier estimates of the quantities that define the MARCH direction, making the update direction fluctuate more strongly from step to step. Increasing the batch size reduces this sampling noise and improves the reliability of the MARCH update.

The comparison shows that \(B\) is not merely a computational-cost parameter. It directly controls the statistical quality of the quantities used to construct the MARCH direction, and therefore affects both the stability and the final accuracy of the optimization.

\subsection{The same training strategy remains applicable across different physical settings}

After identifying the main optimization trends in the baseline doped \(4\times4\) PBC system, we further test whether the same training strategy can be applied beyond this specific setting. The purpose of this section is to examine the applicability of the overall workflow across different fillings, interaction strengths, boundary conditions, and lattice sizes. In all cases, we use the same multi-stage strategy: an NNB initialization stage, supervised Transformer pre-training, and the subsequent MARCH-based VMC optimization.

\subsubsection{Half-filled \(4\times4\) PBC benchmark}

We first consider the half-filled system as a reference case away from the doped baseline. As shown in Fig.~\ref{fig:physical_settings_4x4}(a), the optimization remains stable under the same training protocol, indicating that the method is not restricted to doped systems. The comparison with the doped baseline also shows that the half-filled case is optimized more accurately, consistent with the lower difficulty of this regime.

\subsubsection{Weaker-interaction doped \(4\times4\) PBC system}

We next reduce the interaction strength while keeping the same doped \(4\times4\) PBC geometry, as shown in Fig.~\ref{fig:physical_settings_4x4}(b). Compared with the strongly correlated \(U=8\) doped baseline, the \(U=4\) calculation reaches higher accuracy under the same overall workflow. This behavior is consistent with the reduced correlation strength: the weaker-interaction doped system remains nontrivial, but is less difficult to optimize than the strongly correlated doped case.

\subsubsection{Open-boundary \(4\times4\) systems}

We then test the same training strategy under OBC. This changes the boundary geometry while leaving the optimization protocol unchanged. The results in Fig.~\ref{fig:physical_settings_4x4}(c, d) show that the workflow remains stable for both half-filled and doped OBC systems, suggesting that the method is not tied to PBC.

\subsubsection{Larger \(8\times8\) doped PBC lattice}

Finally, we apply the same overall workflow to a larger doped lattice. This test probes whether the approach can be used beyond the \(4\times4\) benchmark systems. As shown in Fig.~\ref{fig:l8_pbc}, the optimization remains stable on the larger lattice, although the problem is visibly more challenging than the smaller systems.

\section{Discussion}

The results above show that the optimization performance of the Transformer NQS is controlled by three coupled factors: the stability of the MARCH update, the expressive power of the variational ansatz, and the statistical quality of the Monte Carlo estimates. The update-norm threshold \(c_0\) controls how strongly the MARCH direction is damped. A threshold that is too restrictive makes the optimization overly conservative, while a moderate threshold allows the parameters to follow the MARCH direction more effectively. The Transformer width \(d\) and the number of determinant channels \(K\) control two different aspects of expressivity: \(d\) determines the capacity of the hidden representation used to encode nonlocal correlations, while \(K\) determines how many configuration-dependent fermionic structures can be combined in the final wave function. The Monte Carlo batch size \(B\), on the other hand, controls the noise level of the quantities entering the MARCH update.

These factors should not be viewed as independent tuning knobs. Increasing \(d\) or \(K\) enlarges the representational capacity of the ansatz, but this additional capacity can only be useful if the optimization can reliably explore the enlarged parameter space. In practice, this requires both a sufficiently stable MARCH update and sufficiently accurate Monte Carlo estimates. If the norm threshold is too small, the update direction is strongly suppressed even when the ansatz is expressive enough. If the batch size is too small, the stochastic estimates entering the MARCH update become noisy, leading to less stable optimization trajectories. Thus, the improvement of the Transformer NQS is not determined by network size alone, but by the balance between ansatz expressivity, update stability, and sampling quality.

The tests beyond the baseline setting further show that the same workflow is applicable across different physical regimes. The half-filled and weaker-interaction systems are optimized more accurately than the strongly correlated doped baseline, consistent with their lower optimization difficulty. At the same time, the method remains stable for doped systems, OBC, and the larger \(8\times8\) lattice. These results indicate that the combination of NNB initialization, supervised Transformer pre-training, and MARCH-based VMC optimization is not tied to a single lattice size, boundary condition, filling, or interaction strength. Instead, it provides a transferable optimization framework for Transformer-based fermionic NQS calculations.

The present study should be understood as an analysis of optimization trends rather than an exhaustive search over all possible training choices. In particular, the NNB optimization and supervised pre-training schedules are used as fixed initialization procedures within each comparison, and are not treated as independent ablation variables. We also do not separately analyze the role of explicitly imposed symmetry constraints. Therefore, the trends reported here should be interpreted as properties of the current Transformer backflow ansatz and MARCH training protocol without an additional symmetry-enforcement mechanism. A natural next step is to study how symmetry constraints, initialization schedules, model capacity, and MARCH hyperparameters interact with one another in controlling the optimization dynamics.

The present calculations were constrained by the available
computational resources, which consisted of four NVIDIA
GeForce RTX 4090 GPUs. Nevertheless, the comparatively
favorable computational scaling of the backflow Transformer
NQS, approximately $\mathcal{O}(N^3)$ with the number of
lattice sites, suggests that more accurate simulations of
larger and physically relevant lattices should become feasible
as greater computational resources become available.

\begin{acknowledgments}
This work was supported by the National Key R\&D Program of China (Grants No. 2024YFA1408602 and No. 2024YFA1408601) and the National Natural Science Foundation of China (Grant No. 12434009). Zhong-Yi Lu was also supported by the Innovation Program for Quantum Science and Technology (Grant No. 2021ZD0302402). Computational resources were provided by the Physical Laboratory of High Performance Computing in Renmin University of China.
\end{acknowledgments}

\bibliography {main}

@ARTICLE{2025arXiv250702644G,
    author  = {Gu, Yuntian and Li, Wenrui and Lin, Heng and Zhan, Bo
               and Li, Ruichen and Huang, Yifei and He, Di and Wu, Yantao
               and Xiang, Tao and Qin, Mingpu and Wang, Liwei and Lv, Dingshun},
    title   = {Solving the {Hubbard} model with neural quantum states},
    journal = {Nature Communications},
    year    = {2026},
    month   = jun,
    doi     = {10.1038/s41467-026-74028-6}
}

@article{10.1063/1.2746035,
    author  = {Sorella, Sandro and Casula, Michele and Rocca, Dario},
    title   = {Weak binding between two aromatic rings: Feeling the van der Waals attraction by quantum Monte Carlo methods},
    journal = {J. Chem. Phys.},
    volume  = {127},
    number  = {1},
    pages   = {014105},
    year    = {2007},
    doi     = {10.1063/1.2746035}
}

@article{kingma2014adam,
  title={Adam: A method for stochastic optimization},
  author={Kingma, Diederik P and Ba, Jimmy},
  journal={arXiv preprint arXiv:1412.6980},
  year={2014}
}

@article{r4q9-4yvj,
  title = {Accurate Simulation of the Hubbard Model with Finite Fermionic Projected Entangled Pair States},
  author = {Liu, Wen-Yuan and Zhai, Huanchen and Peng, Ruojing and Gu, Zheng-Cheng and Chan, Garnet Kin-Lic},
  journal = {Phys. Rev. Lett.},
  volume = {134},
  issue = {25},
  pages = {256502},
  numpages = {8},
  year = {2025},
  month = {Jun},
  publisher = {American Physical Society},
  doi = {10.1103/r4q9-4yvj},
  url = {https://link.aps.org/doi/10.1103/r4q9-4yvj}
}

@article{PhysRevLett.87.043401,
  title = {Optimization of Ground- and Excited-State Wave Functions and van der Waals Clusters},
  author = {Nightingale, M. P. and Melik-Alaverdian, Vilen},
  journal = {Phys. Rev. Lett.},
  volume = {87},
  issue = {4},
  pages = {043401},
  numpages = {4},
  year = {2001},
  month = {Jul},
  publisher = {American Physical Society},
  doi = {10.1103/PhysRevLett.87.043401},
  url = {https://link.aps.org/doi/10.1103/PhysRevLett.87.043401}
}

@article{PhysRevB.64.024512,
  title = {Generalized Lanczos algorithm for variational quantum Monte Carlo},
  author = {Sorella, Sandro},
  journal = {Phys. Rev. B},
  volume = {64},
  issue = {2},
  pages = {024512},
  numpages = {16},
  year = {2001},
  month = {Jun},
  publisher = {American Physical Society},
  doi = {10.1103/PhysRevB.64.024512},
  url = {https://link.aps.org/doi/10.1103/PhysRevB.64.024512}
}

@article{annurev:/content/journals/10.1146/annurev-conmatphys-090921-033948,
   author = "Qin, Mingpu and Schäfer, Thomas and Andergassen, Sabine and Corboz, Philippe and Gull, Emanuel",
   title = "The Hubbard Model: A Computational Perspective", 
   journal= "Annual Review of Condensed Matter Physics",
   year = "2022",
   volume = "13",
   number = "Volume 13, 2022",
   pages = "275-302",
   doi = "https://doi.org/10.1146/annurev-conmatphys-090921-033948",
   url = "https://www.annualreviews.org/content/journals/10.1146/annurev-conmatphys-090921-033948",
   publisher = "Annual Reviews",
   issn = "1947-5462",
   type = "Journal Article",
  }

@article{annurev:/content/journals/10.1146/annurev-conmatphys-031620-102024,
   author = "Arovas, Daniel P. and Berg, Erez and Kivelson, Steven A. and Raghu, Srinivas",
   title = "The Hubbard Model", 
   journal= "Annual Review of Condensed Matter Physics",
   year = "2022",
   volume = "13",
   number = "Volume 13, 2022",
   pages = "239-274",
   doi = "https://doi.org/10.1146/annurev-conmatphys-031620-102024",
   url = "https://www.annualreviews.org/content/journals/10.1146/annurev-conmatphys-031620-102024",
   publisher = "Annual Reviews",
   issn = "1947-5462",
   type = "Journal Article",
  }

@article{RevModPhys.70.1039,
  title = {Metal-insulator transitions},
  author = {Imada, Masatoshi and Fujimori, Atsushi and Tokura, Yoshinori},
  journal = {Rev. Mod. Phys.},
  volume = {70},
  issue = {4},
  pages = {1039--1263},
  numpages = {0},
  year = {1998},
  month = {Oct},
  publisher = {American Physical Society},
  doi = {10.1103/RevModPhys.70.1039},
  url = {https://link.aps.org/doi/10.1103/RevModPhys.70.1039}
}

@article{PhysRevLett.108.076401,
  title = {Mott Transition in the Two-Dimensional Hubbard Model},
  author = {Kohno, Masanori},
  journal = {Phys. Rev. Lett.},
  volume = {108},
  issue = {7},
  pages = {076401},
  numpages = {5},
  year = {2012},
  month = {Feb},
  publisher = {American Physical Society},
  doi = {10.1103/PhysRevLett.108.076401},
  url = {https://link.aps.org/doi/10.1103/PhysRevLett.108.076401}
}

@article{PhysRevB.95.235109,
  title = {Signatures of the Mott transition in the antiferromagnetic state of the two-dimensional Hubbard model},
  author = {Fratino, L. and S\'emon, P. and Charlebois, M. and Sordi, G. and Tremblay, A.-M. S.},
  journal = {Phys. Rev. B},
  volume = {95},
  issue = {23},
  pages = {235109},
  numpages = {11},
  year = {2017},
  month = {Jun},
  publisher = {American Physical Society},
  doi = {10.1103/PhysRevB.95.235109},
  url = {https://link.aps.org/doi/10.1103/PhysRevB.95.235109}
}

@article{RevModPhys.68.13,
  title = {Dynamical mean-field theory of strongly correlated fermion systems and the limit of infinite dimensions},
  author = {Georges, Antoine and Kotliar, Gabriel and Krauth, Werner and Rozenberg, Marcelo J.},
  journal = {Rev. Mod. Phys.},
  volume = {68},
  issue = {1},
  pages = {13--125},
  numpages = {0},
  year = {1996},
  month = {Jan},
  publisher = {American Physical Society},
  doi = {10.1103/RevModPhys.68.13},
  url = {https://link.aps.org/doi/10.1103/RevModPhys.68.13}
}

@article{PhysRevX.11.041013,
  title = {Mott Insulating States with Competing Orders in the Triangular Lattice Hubbard Model},
  author = {Wietek, Alexander and Rossi, Riccardo and \ifmmode \check{S}\else \v{S}\fi{}imkovic, Fedor and Klett, Marcel and Hansmann, Philipp and Ferrero, Michel and Stoudenmire, E. Miles and Sch\"afer, Thomas and Georges, Antoine},
  journal = {Phys. Rev. X},
  volume = {11},
  issue = {4},
  pages = {041013},
  numpages = {23},
  year = {2021},
  month = {Oct},
  publisher = {American Physical Society},
  doi = {10.1103/PhysRevX.11.041013},
  url = {https://link.aps.org/doi/10.1103/PhysRevX.11.041013}
}

@article{Mazurenko2017FermiHubbardAF, title = {A cold-atom Fermi-Hubbard antiferromagnet}, author = {Mazurenko, Anton and Chiu, Christie S. and Ji, Geoffrey and Parsons, Maxwell F. and Kan{\'a}sz-Nagy, M{\'a}rton and Schmidt, Richard and Grusdt, Fabian and Demler, Eugene and Greif, Daniel and Greiner, Markus}, journal = {Nature}, volume = {545}, pages = {462--466}, year = {2017}, doi = {10.1038/nature22362} }

@article{Cheuk2016SpinCharge2DFH, title = {Observation of spatial charge and spin correlations in the 2D Fermi-Hubbard model}, author = {Cheuk, Lawrence W. and Nichols, Matthew A. and Lawrence, Katherine R. and Okan, Melih and Zhang, Hao and Khatami, Ehsan and Trivedi, Nandini and Paiva, Thereza and Rigol, Marcos and Zwierlein, Martin W.}, journal = {Science}, volume = {353}, number = {6305}, pages = {1260--1264}, year = {2016}, doi = {10.1126/science.aag3349} }

@article{Parsons2016SpinCorrelationFH, title = {Site-resolved measurement of the spin-correlation function in the Fermi-Hubbard model}, author = {Parsons, Maxwell F. and Mazurenko, Anton and Chiu, Christie S. and Ji, Geoffrey and Greif, Daniel and Greiner, Markus}, journal = {Science}, volume = {353}, number = {6305}, pages = {1253--1256}, year = {2016}, doi = {10.1126/science.aag1430} }

@article{Zheng2017StripeOrder, title = {Stripe order in the underdoped region of the two-dimensional Hubbard model}, author = {Zheng, Bo-Xiao and Chung, Chia-Min and Corboz, Philippe and Ehlers, Georg and Qin, Ming-Pu and Noack, Reinhard M. and Shi, Hao and White, Steven R. and Zhang, Shiwei and Chan, Garnet Kin-Lic}, journal = {Science}, volume = {358}, number = {6367}, pages = {1155--1160}, year = {2017}, doi = {10.1126/science.aam7127}}

@article{PhysRevResearch.4.013239,
  title = {Stripes and spin-density waves in the doped two-dimensional Hubbard model: Ground state phase diagram},
  author = {Xu, Hao and Shi, Hao and Vitali, Ettore and Qin, Mingpu and Zhang, Shiwei},
  journal = {Phys. Rev. Res.},
  volume = {4},
  issue = {1},
  pages = {013239},
  numpages = {10},
  year = {2022},
  month = {Mar},
  publisher = {American Physical Society},
  doi = {10.1103/PhysRevResearch.4.013239},
  url = {https://link.aps.org/doi/10.1103/PhysRevResearch.4.013239}
}

@article{PhysRevB.80.075116,
  title = {Quantum Monte Carlo study of the two-dimensional fermion Hubbard model},
  author = {Varney, C. N. and Lee, C.-R. and Bai, Z. J. and Chiesa, S. and Jarrell, M. and Scalettar, R. T.},
  journal = {Phys. Rev. B},
  volume = {80},
  issue = {7},
  pages = {075116},
  numpages = {8},
  year = {2009},
  month = {Aug},
  publisher = {American Physical Society},
  doi = {10.1103/PhysRevB.80.075116},
  url = {https://link.aps.org/doi/10.1103/PhysRevB.80.075116}
}

@article{PhysRevLett.124.017003,
  title = {Extended Crossover from a Fermi Liquid to a Quasiantiferromagnet in the Half-Filled 2D Hubbard Model},
  author = {\ifmmode \check{S}\else \v{S}\fi{}imkovic, Fedor and LeBlanc, J. P. F. and Kim, Aaram J. and Deng, Youjin and Prokof'ev, N. V. and Svistunov, B. V. and Kozik, Evgeny},
  journal = {Phys. Rev. Lett.},
  volume = {124},
  issue = {1},
  pages = {017003},
  numpages = {6},
  year = {2020},
  month = {Jan},
  publisher = {American Physical Society},
  doi = {10.1103/PhysRevLett.124.017003},
  url = {https://link.aps.org/doi/10.1103/PhysRevLett.124.017003}
}

@article{PhysRevB.69.085119,
  title = {Antiferromagnetism and single-particle properties in the two-dimensional half-filled Hubbard model: A nonlinear sigma model approach},
  author = {Borejsza, K. and Dupuis, N.},
  journal = {Phys. Rev. B},
  volume = {69},
  issue = {8},
  pages = {085119},
  numpages = {17},
  year = {2004},
  month = {Feb},
  publisher = {American Physical Society},
  doi = {10.1103/PhysRevB.69.085119},
  url = {https://link.aps.org/doi/10.1103/PhysRevB.69.085119}
}

@article{PhysRevX.13.011007,
  title = {Temperature Dependence of Spin and Charge Orders in the Doped Two-Dimensional Hubbard Model},
  author = {Xiao, Bo and He, Yuan-Yao and Georges, Antoine and Zhang, Shiwei},
  journal = {Phys. Rev. X},
  volume = {13},
  issue = {1},
  pages = {011007},
  numpages = {20},
  year = {2023},
  month = {Jan},
  publisher = {American Physical Society},
  doi = {10.1103/PhysRevX.13.011007},
  url = {https://link.aps.org/doi/10.1103/PhysRevX.13.011007}
}

@article{PhysRevLett.104.116402,
  title = {Spin and Charge Order in the Doped Hubbard Model: Long-Wavelength Collective Modes},
  author = {Chang, Chia-Chen and Zhang, Shiwei},
  journal = {Phys. Rev. Lett.},
  volume = {104},
  issue = {11},
  pages = {116402},
  numpages = {4},
  year = {2010},
  month = {Mar},
  publisher = {American Physical Society},
  doi = {10.1103/PhysRevLett.104.116402},
  url = {https://link.aps.org/doi/10.1103/PhysRevLett.104.116402}
}

@article{Mai2022IntertwinedSpinChargePair,
  title   = {Intertwined spin, charge, and pair correlations in the two-dimensional Hubbard model in the thermodynamic limit},
  author  = {Mai, Peizhi and Karakuzu, Seher and Balduzzi, Giovanni and Johnston, Steven and Maier, Thomas A.},
  journal = {Proceedings of the National Academy of Sciences},
  volume  = {119},
  number  = {7},
  pages   = {e2112806119},
  year    = {2022},
  doi     = {10.1073/pnas.2112806119}
}

@article{Huang2018StripeOrderHubbard,
  title   = {Stripe order from the perspective of the Hubbard model},
  author  = {Huang, Edwin W. and Mendl, Christian B. and Jiang, Hong-Chen and Moritz, Brian and Devereaux, Thomas P.},
  journal = {npj Quantum Materials},
  volume  = {3},
  pages   = {22},
  year    = {2018},
  doi     = {10.1038/s41535-018-0097-0}
}

@article{Jiang2019DopedHubbardSC,
  title   = {Superconductivity in the doped Hubbard model and its interplay with next-nearest hopping \(t'\)},
  author  = {Jiang, Hong-Chen and Devereaux, Thomas P.},
  journal = {Science},
  volume  = {365},
  number  = {6460},
  pages   = {1424--1428},
  year    = {2019},
  doi     = {10.1126/science.aal5304}
}

@article{PhysRevX.10.031016,
  title = {Absence of Superconductivity in the Pure Two-Dimensional Hubbard Model},
  author = {Qin, Mingpu and Chung, Chia-Min and Shi, Hao and Vitali, Ettore and Hubig, Claudius and Schollw\"ock, Ulrich and White, Steven R. and Zhang, Shiwei},
  collaboration = {Simons Collaboration on the Many-Electron Problem},
  journal = {Phys. Rev. X},
  volume = {10},
  issue = {3},
  pages = {031016},
  numpages = {18},
  year = {2020},
  month = {Jul},
  publisher = {American Physical Society},
  doi = {10.1103/PhysRevX.10.031016},
  url = {https://link.aps.org/doi/10.1103/PhysRevX.10.031016}
}

@article{PhysRevB.109.085121,
  title = {Ground-state phase diagram and superconductivity of the doped Hubbard model on six-leg square cylinders},
  author = {Jiang, Yi-Fan and Devereaux, Thomas P. and Jiang, Hong-Chen},
  journal = {Phys. Rev. B},
  volume = {109},
  issue = {8},
  pages = {085121},
  numpages = {6},
  year = {2024},
  month = {Feb},
  publisher = {American Physical Society},
  doi = {10.1103/PhysRevB.109.085121},
  url = {https://link.aps.org/doi/10.1103/PhysRevB.109.085121}
}

@misc{scalapino2006numericalstudies2dhubbard,
      title={Numerical Studies of the 2D Hubbard Model}, 
      author={D. J. Scalapino},
      year={2006},
      eprint={cond-mat/0610710},
      archivePrefix={arXiv},
      primaryClass={cond-mat.str-el},
      url={https://arxiv.org/abs/cond-mat/0610710}, 
}

@article{Dong2022MechanismHubbardSC,
  title   = {Mechanism of superconductivity in the Hubbard model at intermediate interaction strength},
  author  = {Dong, Xinyang and Del Re, Lorenzo and Toschi, Alessandro and Gull, Emanuel},
  journal = {Proceedings of the National Academy of Sciences},
  volume  = {119},
  number  = {33},
  pages   = {e2205048119},
  year    = {2022},
  doi     = {10.1073/pnas.2205048119}
}

@article{PhysRevB.71.134527,
  title = {Physics of cuprates with the two-band Hubbard model: The validity of the one-band Hubbard model},
  author = {Macridin, A. and Jarrell, M. and Maier, Th. and Sawatzky, G. A.},
  journal = {Phys. Rev. B},
  volume = {71},
  issue = {13},
  pages = {134527},
  numpages = {13},
  year = {2005},
  month = {Apr},
  publisher = {American Physical Society},
  doi = {10.1103/PhysRevB.71.134527},
  url = {https://link.aps.org/doi/10.1103/PhysRevB.71.134527}
}

@article{PhysRevLett.85.1524,
  title = {$\mathit{d}$-Wave Superconductivity in the Hubbard Model},
  author = {Maier, Th. and Jarrell, M. and Pruschke, Th. and Keller, J.},
  journal = {Phys. Rev. Lett.},
  volume = {85},
  issue = {7},
  pages = {1524--1527},
  numpages = {0},
  year = {2000},
  month = {Aug},
  publisher = {American Physical Society},
  doi = {10.1103/PhysRevLett.85.1524},
  url = {https://link.aps.org/doi/10.1103/PhysRevLett.85.1524}
}

@article{PhysRevLett.110.216405,
  title = {Superconductivity and the Pseudogap in the Two-Dimensional Hubbard Model},
  author = {Gull, Emanuel and Parcollet, Olivier and Millis, Andrew J.},
  journal = {Phys. Rev. Lett.},
  volume = {110},
  issue = {21},
  pages = {216405},
  numpages = {5},
  year = {2013},
  month = {May},
  publisher = {American Physical Society},
  doi = {10.1103/PhysRevLett.110.216405},
  url = {https://link.aps.org/doi/10.1103/PhysRevLett.110.216405}
}

@article{PhysRevB.55.7464,
  title = {Constrained path Monte Carlo method for fermion ground states},
  author = {Zhang, Shiwei and Carlson, J. and Gubernatis, J. E.},
  journal = {Phys. Rev. B},
  volume = {55},
  issue = {12},
  pages = {7464--7477},
  numpages = {0},
  year = {1997},
  month = {Mar},
  publisher = {American Physical Society},
  doi = {10.1103/PhysRevB.55.7464},
  url = {https://link.aps.org/doi/10.1103/PhysRevB.55.7464}
}

@article{PhysRevB.94.085103,
  title = {Benchmark study of the two-dimensional Hubbard model with auxiliary-field quantum Monte Carlo method},
  author = {Qin, Mingpu and Shi, Hao and Zhang, Shiwei},
  journal = {Phys. Rev. B},
  volume = {94},
  issue = {8},
  pages = {085103},
  numpages = {12},
  year = {2016},
  month = {Aug},
  publisher = {American Physical Society},
  doi = {10.1103/PhysRevB.94.085103},
  url = {https://link.aps.org/doi/10.1103/PhysRevB.94.085103}
}

@article{PhysRevB.96.075156,
  title = {Numerical results on the short-range spin correlation functions in the ground state of the two-dimensional Hubbard model},
  author = {Qin, Mingpu and Shi, Hao and Zhang, Shiwei},
  journal = {Phys. Rev. B},
  volume = {96},
  issue = {7},
  pages = {075156},
  numpages = {7},
  year = {2017},
  month = {Aug},
  publisher = {American Physical Society},
  doi = {10.1103/PhysRevB.96.075156},
  url = {https://link.aps.org/doi/10.1103/PhysRevB.96.075156}
}

@article{PhysRevB.107.235124,
  title = {Self-consistent optimization of the trial wave function within the constrained path auxiliary field quantum Monte Carlo method using mixed estimators},
  author = {Qin, Mingpu},
  journal = {Phys. Rev. B},
  volume = {107},
  issue = {23},
  pages = {235124},
  numpages = {6},
  year = {2023},
  month = {Jun},
  publisher = {American Physical Society},
  doi = {10.1103/PhysRevB.107.235124},
  url = {https://link.aps.org/doi/10.1103/PhysRevB.107.235124}
}

@article{PhysRevLett.91.136403,
  title = {Stripes on a 6-Leg Hubbard Ladder},
  author = {White, Steven R. and Scalapino, D. J.},
  journal = {Phys. Rev. Lett.},
  volume = {91},
  issue = {13},
  pages = {136403},
  numpages = {4},
  year = {2003},
  month = {Sep},
  publisher = {American Physical Society},
  doi = {10.1103/PhysRevLett.91.136403},
  url = {https://link.aps.org/doi/10.1103/PhysRevLett.91.136403}
}

@article{PhysRevB.71.075108,
  title = {Stripe formation in doped Hubbard ladders},
  author = {Hager, G. and Wellein, G. and Jeckelmann, E. and Fehske, H.},
  journal = {Phys. Rev. B},
  volume = {71},
  issue = {7},
  pages = {075108},
  numpages = {6},
  year = {2005},
  month = {Feb},
  publisher = {American Physical Society},
  doi = {10.1103/PhysRevB.71.075108},
  url = {https://link.aps.org/doi/10.1103/PhysRevB.71.075108}
}

@article{PhysRevB.95.125125,
  title = {Hybrid-space density matrix renormalization group study of the doped two-dimensional Hubbard model},
  author = {Ehlers, G. and White, S. R. and Noack, R. M.},
  journal = {Phys. Rev. B},
  volume = {95},
  issue = {12},
  pages = {125125},
  numpages = {14},
  year = {2017},
  month = {Mar},
  publisher = {American Physical Society},
  doi = {10.1103/PhysRevB.95.125125},
  url = {https://link.aps.org/doi/10.1103/PhysRevB.95.125125}
}

@article{PhysRevB.93.045116,
  title = {Improved energy extrapolation with infinite projected entangled-pair states applied to the two-dimensional Hubbard model},
  author = {Corboz, Philippe},
  journal = {Phys. Rev. B},
  volume = {93},
  issue = {4},
  pages = {045116},
  numpages = {7},
  year = {2016},
  month = {Jan},
  publisher = {American Physical Society},
  doi = {10.1103/PhysRevB.93.045116},
  url = {https://link.aps.org/doi/10.1103/PhysRevB.93.045116}
}

@article{PhysRevB.107.165112,
  title = {Finite projected entangled pair states for the Hubbard model},
  author = {Scheb, M. and Noack, R. M.},
  journal = {Phys. Rev. B},
  volume = {107},
  issue = {16},
  pages = {165112},
  numpages = {28},
  year = {2023},
  month = {Apr},
  publisher = {American Physical Society},
  doi = {10.1103/PhysRevB.107.165112},
  url = {https://link.aps.org/doi/10.1103/PhysRevB.107.165112}
}

@article{PhysRevB.100.195141,
  title = {Period 4 stripe in the extended two-dimensional Hubbard model},
  author = {Ponsioen, Boris and Chung, Sangwoo S. and Corboz, Philippe},
  journal = {Phys. Rev. B},
  volume = {100},
  issue = {19},
  pages = {195141},
  numpages = {7},
  year = {2019},
  month = {Nov},
  publisher = {American Physical Society},
  doi = {10.1103/PhysRevB.100.195141},
  url = {https://link.aps.org/doi/10.1103/PhysRevB.100.195141}
}

@article{PhysRevLett.134.116502,
  title = {Frustration-Induced Superconductivity in the $t\text{\ensuremath{-}}{t}^{\ensuremath{'}}$ Hubbard Model},
  author = {Zhang, Changkai and Li, Jheng-Wei and Nikolaidou, Dimitra and von Delft, Jan},
  journal = {Phys. Rev. Lett.},
  volume = {134},
  issue = {11},
  pages = {116502},
  numpages = {7},
  year = {2025},
  month = {Mar},
  publisher = {American Physical Society},
  doi = {10.1103/PhysRevLett.134.116502},
  url = {https://link.aps.org/doi/10.1103/PhysRevLett.134.116502}
}

@article{PhysRevLett.109.186404,
  title = {Density Matrix Embedding: A Simple Alternative to Dynamical Mean-Field Theory},
  author = {Knizia, Gerald and Chan, Garnet Kin-Lic},
  journal = {Phys. Rev. Lett.},
  volume = {109},
  issue = {18},
  pages = {186404},
  numpages = {5},
  year = {2012},
  month = {Nov},
  publisher = {American Physical Society},
  doi = {10.1103/PhysRevLett.109.186404},
  url = {https://link.aps.org/doi/10.1103/PhysRevLett.109.186404}
}

@article{PhysRevB.93.035126,
  title = {Ground-state phase diagram of the square lattice Hubbard model from density matrix embedding theory},
  author = {Zheng, Bo-Xiao and Chan, Garnet Kin-Lic},
  journal = {Phys. Rev. B},
  volume = {93},
  issue = {3},
  pages = {035126},
  numpages = {17},
  year = {2016},
  month = {Jan},
  publisher = {American Physical Society},
  doi = {10.1103/PhysRevB.93.035126},
  url = {https://link.aps.org/doi/10.1103/PhysRevB.93.035126}
}

@article{Wu2019ProjectedDMET,
  title   = {Projected density matrix embedding theory with applications to the two-dimensional Hubbard model},
  author  = {Wu, Xiaojie and Cui, Zhi-Hao and Tong, Yu and Lindsey, Michael and Chan, Garnet Kin-Lic and Lin, Lin},
  journal = {Journal of Chemical Physics},
  volume  = {151},
  pages   = {064108},
  year    = {2019},
  doi     = {10.1063/1.5108818}
}

@article{Carleo2017NQS,
  title   = {Solving the Quantum Many-Body Problem with Artificial Neural Networks},
  author  = {Carleo, Giuseppe and Troyer, Matthias},
  journal = {Science},
  volume  = {355},
  number  = {6325},
  pages   = {602--606},
  year    = {2017},
  doi     = {10.1126/science.aag2302}
}

@article{PhysRevB.96.205152,
  title = {Restricted Boltzmann machine learning for solving strongly correlated quantum systems},
  author = {Nomura, Yusuke and Darmawan, Andrew S. and Yamaji, Youhei and Imada, Masatoshi},
  journal = {Phys. Rev. B},
  volume = {96},
  issue = {20},
  pages = {205152},
  numpages = {8},
  year = {2017},
  month = {Nov},
  publisher = {American Physical Society},
  doi = {10.1103/PhysRevB.96.205152},
  url = {https://link.aps.org/doi/10.1103/PhysRevB.96.205152}
}

@article{PhysRevX.8.011006,
  title = {Neural-Network Quantum States, String-Bond States, and Chiral Topological States},
  author = {Glasser, Ivan and Pancotti, Nicola and August, Moritz and Rodriguez, Ivan D. and Cirac, J. Ignacio},
  journal = {Phys. Rev. X},
  volume = {8},
  issue = {1},
  pages = {011006},
  numpages = {16},
  year = {2018},
  month = {Jan},
  publisher = {American Physical Society},
  doi = {10.1103/PhysRevX.8.011006},
  url = {https://link.aps.org/doi/10.1103/PhysRevX.8.011006}
}

@article{PhysRevB.109.245120,
  title = {Variational optimization of the amplitude of neural-network quantum many-body ground states},
  author = {Wang, Jia-Qi and Wu, Han-Qing and He, Rong-Qiang and Lu, Zhong-Yi},
  journal = {Phys. Rev. B},
  volume = {109},
  issue = {24},
  pages = {245120},
  numpages = {8},
  year = {2024},
  month = {Jun},
  publisher = {American Physical Society},
  doi = {10.1103/PhysRevB.109.245120},
  url = {https://link.aps.org/doi/10.1103/PhysRevB.109.245120}
}

@article{m4c7-qz8l,
  title = {Generalized Lanczos method for systematic optimization of neural-network quantum states},
  author = {Wang, Jia-Qi and He, Rong-Qiang and Lu, Zhong-Yi},
  journal = {Phys. Rev. B},
  volume = {113},
  issue = {8},
  pages = {085120},
  numpages = {12},
  year = {2026},
  month = {Feb},
  publisher = {American Physical Society},
  doi = {10.1103/m4c7-qz8l},
  url = {https://link.aps.org/doi/10.1103/m4c7-qz8l}
}

@article{Luo2019Backflow,
  author  = {Luo, Di and Clark, Bryan K.},
  title   = {Backflow Transformations via Neural Networks for Quantum Many-Body Wave Functions},
  journal = {Physical Review Letters},
  volume  = {122},
  number  = {22},
  pages   = {226401},
  year    = {2019},
  doi     = {10.1103/PhysRevLett.122.226401}
}

@article{Elfwing2018SiLU,
  author  = {Elfwing, Stefan and Uchibe, Eiji and Doya, Kenji},
  title   = {Sigmoid-Weighted Linear Units for Neural Network
             Function Approximation in Reinforcement Learning},
  journal = {Neural Networks},
  volume  = {107},
  pages   = {3--11},
  year    = {2018},
  doi     = {10.1016/j.neunet.2018.04.020}
}

@article{Pfau2020FermiNet,
  author  = {Pfau, David and Spencer, James S. and
             Matthews, Alexander G. D. G. and
             Foulkes, W. M. C.},
  title   = {Ab Initio Solution of the Many-Electron
             Schr{\"o}dinger Equation with Deep Neural Networks},
  journal = {Physical Review Research},
  volume  = {2},
  number  = {3},
  pages   = {033429},
  year    = {2020},
  doi     = {10.1103/PhysRevResearch.2.033429}
}

@article{Chen2024EfficientOptimization,
  author  = {Chen, Ao and Heyl, Markus},
  title   = {Empowering deep neural quantum states through efficient optimization},
  journal = {Nature Physics},
  volume  = {20},
  pages   = {1476--1481},
  year    = {2024},
  doi     = {10.1038/s41567-024-02566-1},
  url     = {https://doi.org/10.1038/s41567-024-02566-1}
}

@article{Rende2024LargeScaleSR,
  author  = {Rende, Riccardo and Viteritti, Luciano Loris and Bardone, Lorenzo and Becca, Federico and Goldt, Sebastian},
  title   = {A simple linear algebra identity to optimize large-scale neural network quantum states},
  journal = {Communications Physics},
  volume  = {7},
  pages   = {260},
  year    = {2024},
  doi     = {10.1038/s42005-024-01732-4},
  url     = {https://doi.org/10.1038/s42005-024-01732-4}
}

@article{RobledoMoreno2022HiddenFermion,
  author  = {Robledo Moreno, Javier and Carleo, Giuseppe and Georges, Antoine and Stokes, James},
  title   = {Fermionic wave functions from neural-network constrained hidden states},
  journal = {Proceedings of the National Academy of Sciences},
  volume  = {119},
  number  = {32},
  pages   = {e2122059119},
  year    = {2022},
  doi     = {10.1073/pnas.2122059119},
  url     = {https://doi.org/10.1073/pnas.2122059119}
}

@misc{Chen2025PfaffianNQS,
  author        = {Chen, Ao and Wan, Zhou-Quan and Sengupta, Anirvan and Georges, Antoine and Roth, Christopher},
  title         = {Neural Network-Augmented Pfaffian Wave-functions for Scalable Simulations of Interacting Fermions},
  year          = {2025},
  eprint        = {2507.10705},
  archivePrefix = {arXiv},
  primaryClass  = {cond-mat.str-el},
  doi           = {10.48550/arXiv.2507.10705},
  url           = {https://arxiv.org/abs/2507.10705}
}

@article{IbarraGarciaPadilla2025AutoNQS,
  author  = {Ibarra-Garc{\'i}a-Padilla, Eduardo and Lange, Hannah and Melko, Roger G. and Scalettar, Richard T. and Carrasquilla, Juan and Bohrdt, Annabelle and Khatami, Ehsan},
  title   = {Autoregressive neural quantum states of Fermi Hubbard models},
  journal = {Physical Review Research},
  volume  = {7},
  pages   = {013122},
  year    = {2025},
  doi     = {10.1103/PhysRevResearch.7.013122},
  url     = {https://doi.org/10.1103/PhysRevResearch.7.013122}
}

@misc{Gu2026ParetoBackflow,
  author        = {Gu, Yuntian and Han, Zeyao and Li, Wenrui and Xiao, Zhiyu and Xiang, Tao and Qin, Mingpu and Wang, Liwei and Lv, Dingshun},
  title         = {Pareto Frontier of Neural Quantum States: Scalable, Affordable, and Accurate Convolutional Backflow for Strongly Correlated Lattice Fermions},
  year          = {2026},
  eprint        = {2604.25775},
  archivePrefix = {arXiv},
  primaryClass  = {cond-mat.str-el},
  doi           = {10.48550/arXiv.2604.25775},
  url           = {https://arxiv.org/abs/2604.25775}
}

@misc{Viteritti2026VariationalBias,
  author        = {Viteritti, Luciano Loris and Rende, Riccardo and Roth, Christopher and Sengupta, Anirvan and Carleo, Giuseppe and Georges, Antoine},
  title         = {Beyond Variational Bias: Resolving Intertwined Orders in the Hubbard Model},
  year          = {2026},
  eprint        = {2604.21978},
  archivePrefix = {arXiv},
  primaryClass  = {cond-mat.str-el},
  doi           = {10.48550/arXiv.2604.21978},
  url           = {https://arxiv.org/abs/2604.21978}
}
\end{document}